\documentclass[12pt,preprint]{aastex}

\newcommand{\etal}{et~al.\ }

\newcommand{\be}{\begin{equation}}
\newcommand{\ee}{\end{equation}}
\newcommand{\ba}{\begin{eqnarray}}
\newcommand{\ea}{\end{eqnarray}}

\newcommand{\swift}{\emph{Swift}}

\begin{document}

\def\Mesz{M\'esz\'aros}
\def\sarc{$^{\prime\prime}\!\!.$}
\def\arcsec{$^{\prime\prime}$}
\def\ls{\lower 2pt \hbox{$\;\scriptscriptstyle \buildrel<\over\sim\;$}}
\def\gs{\lower 2pt \hbox{$\;\scriptscriptstyle \buildrel>\over\sim\;$}}

\title{Optical and X-Ray Observations of GRB 060526: A Complex Afterglow Consistent with An Achromatic Jet Break}

\author{X.~Dai\altaffilmark{1}, J.~P.~Halpern\altaffilmark{2}, N.~D.~Morgan\altaffilmark{1}, E.~Armstrong\altaffilmark{3}, N.~Mirabal\altaffilmark{2,4}, J.~B.~Haislip\altaffilmark{5}, D.~E.~Reichart\altaffilmark{5}, and K.~Z.~Stanek\altaffilmark{1}}

\altaffiltext{1}{Department of Astronomy,
Ohio State University, Columbus, OH 43210.}

\altaffiltext{2}{Department of Astronomy, Columbia University, 550 West 120th Street, New York, NY 10027.}

\altaffiltext{3}{Department of Physics, University of California, San Diego, CA 92093.}

\altaffiltext{4}{Department of Astronomy, University of Michigan, Ann Arbor, MI 48109.}

\altaffiltext{5}{Department of Physics and Astronomy, University of North Carolina, Chapel Hill, NC 27599.}

\email{xinyu@astronomy.ohio-state.edu}

\begin{abstract}
We obtained 98 $R$-band and 18 $B$, $r'$, $i'$ images of the optical afterglow of GRB 060526 ($z=3.21$) 
with the MDM 1.3m, 2.4m, and the PROMPT telescopes in Cerro Tololo
 over the 5 nights following the burst trigger.  
Combining these data with other optical observations reported in GCN and the \swift-XRT
observations, we compare the optical and X-ray afterglow light curves of GRB 060526.
Both the optical and X-ray afterglow light curves show rich features, 
such as flares and breaks.  
The densely sampled optical observations provide very good coverage at $T>10^4$~sec.
We observed a break at $2.4\times10^5$~sec in the optical afterglow light curve.
Compared with the X-ray afterglow light curve, the break is consistent with an achromatic break
supporting the beaming models of GRBs.
However, the pre-break and post-break temporal decay slopes are difficult to explain in simple afterglow models.
We estimated a jet angle of $\theta_j \sim 7^{\circ}$ and a prompt emission size of $R_{prompt} \sim 2\times10^{14}$~cm.
In addition, we detected several optical flares with amplitudes of $\Delta m \sim 0.2$, 0.6, and 0.2~mag.  The X-ray afterglows detected by \swift\ have shown complicated decay patterns.  Recently, many well-sampled optical afterglows also show decays with flares and multiple breaks.  
GRB~060526 provides an additional case of such a complex, well observed optical afterglow.  
The accumulated well-sampled afterglows indicate that most of the optical afterglows are complex.

\end{abstract}

\keywords{gamma rays: bursts}

\section{Introduction}

In the fireball model \citep{meszaros97,sari98}, the afterglow emission of gamma-ray bursts (GRBs) are thought to be synchrotron emission in the external shocks. 
After the launch of \swift\ \citep{gehrels04}, with its rapid localization of GRBs and the dedicated on-board XRT instrument, the afterglow models can be tested extensively with the regularly obtained XRT light curves.
More than half of the \swift-XRT light curves show complicated decay patterns with multiple breaks and giant X-ray flares \citep{burrows05,nousek06,obrien06}.  New ingredients were added to the models to interpret these features \citep[e.g.,][]{zhang06,nousek06,panaitescu06a}.  Compared with the large number of \swift-XRT afterglows, only a few bursts have good optical afterglow coverage, which limits the multi-wavelength study of GRB afterglows.  Moreover, a large fraction of the well-studied optical afterglows also show complicated behaviors (e.g., Guidorzi et al. 2005; Blustin et al. 2006; Rykoff et al. 2006; Stanek et al. 2006), challenging the simple, smooth-decay afterglow models.  

Another important aspect of the models is that GRBs are thought to be collimated in jets,
based on the achromatic breaks observed in many optical GRB
afterglow light curves \citep[e.g.,][]{stanek99}.  
Different jet models have been proposed for GRBs either under a uniform jet model \citep[e.g.,][]{rhoads99,frail01,granot02} or a structured jet model \citep[e.g.,][]{lipunov01,zhang02,rossi02,lloyd-ronning04,zhang04,dai05}.  
Some of these jet models can also unify the closely related phenomena of
X-ray flashes with GRBs \citep{yamazaki03,zhang04,lamb05,dai05}.
Since the distinct signature that a jet imposes on GRB afterglow light curves (an achromatic break) is simply a geometric effect, it is important to test the wavelength independence across the broadest possible wavelength range, for example between optical and X-ray afterglow light curves.  To date, the lack of wavelength dependence has only been confirmed across different optical bands.  Recently, GRB~050525A \citep{blustin06}, GRB~050801 \citep{rykoff06a}, and GRB~060206 \citep{stanek06} show possible achromatic breaks across optical and X-ray light curves.  However, in GRB~050801, the break is interpreted as energy injection or the onset of the afterglow, and in GRB~060206, it is debated whether the break is achromatic \citep{monfardini06}.

As many XRT light curves show multiple breaks, it is not obvious which of them should be associated with the optical break and which of them interpreted as the jet break.  
Recently, \citet{panaitescu06b,fan06} showed that some of the X-ray breaks (1--4 hours after the burst trigger) are chromatic from X-rays to optical bands.
However, as the X-ray light curves have several breaks, it is possible that the achromatic jet break occurs at some later time.
In addition, as many optical afterglows also show rich features such as flares and multiple breaks (e.g., Stanek et al. 2006), fits to poorly sampled optical light curves may not be reliable. 

In this paper, we report the optical follow-up of GRB 060526 with the MDM 1.3m, 2.4m telescopes and the PROMPT at Cerro Tololo,
and our detection of an achromatic break across the optical and X-ray bands.  
GRB 060526 was detected by the BAT on board \swift\ at 16:28:30 UT on May 26, 2006 (Campana et al. 2006).
The XRT and UVOT rapidly localized the burst location.  The burst was followed up with ground-based
telescopes by several groups.  In particular, 
\citet{berger06} reported the burst redshift of $z=3.21$.
We organize the paper as follows.  First, we describe the data reduction in \S\ref{sec:data}.  In \S\ref{sec:evo}, we describe the evolution of the GRB afterglow and perform a comparison between the optical and X-ray light curves.  Finally, we discuss our results in \S\ref{sec:dis}.

\section{The Optical and X-ray Data Reduction\label{sec:data}}
We obtained 83 and 15 optical $R$-band exposures with the MDM 1.3m, 2.4m telescopes, 
and 18 $B$, $r'$, $i'$ images with the PROMPT in Cerro Tololo, respectively, on
the nights of 26-30 May 2006.
After standard bias subtraction and flatfielding, we measured
relative fluxes between GRB~060526 and six nearby reference stars that had
been calibrated using the Landolt (1992) standard field on the 1.3m. The local
reference stars also served to tie the 2.4m data onto the 1.3m photometric
system, with typical rms scatter of the 2.4m zeropoint determination of
0.02-0.03 magnitudes for each epoch.  In general, overlapping GRB~060526
data between the two telescopes agreed to within the computed error bars
(see Fig.~\ref{fig:opt}).  
As GRB~060526 was a bright burst observed by many groups, we also collected other R-band 
observations reported in the GCN circulars  
\citep{french06,covino06,lin06,khamitov06a,khamitov06b,khamitov06c,khamitov06d,khamitov06e,khamitov06f,rumyantsev06a,rumyantsev06b,kann06,baliyan06,terra06}.  When the reference stars and their magnitudes were given,
we calibrated their magnitudes to our magnitude system.

We also reduced the \swift-XRT data for GRB 060526.  The XRT observations cover 6 days with gaps in between for the burst.  We analyzed the XRT level 2 event files for both the windowed timing (WT) and photon counting (PC) modes.  These events were filtered to be within 0.2-10 keV energy band and restricted to grades of 0--2 for the WT mode and 0--12 for the PC mode.  We extracted the XRT spectra and light-curves with the software tool \verb+xselect+ \footnote{http://heasarc.gsfc.nasa.gov/docs/software/lheasoft/ftools/xselect/xselect.html.}.  We use the \verb+rmf+ files from the standard XRT calibration distribution, and
generated the \verb+arf+ files with the \emph{Swift}-XRT software tool
 \footnote{http://swift.gsfc.nasa.gov/docs/swift/analysis/xrt\_swguide\_v1\_2.pdf.}.
Finally, we fit the X-ray spectra with \verb+XSPEC+ \citep{arnaud96}.

\section{Evolution of the Afterglow\label{sec:evo}}
\begin{figure}
\plotone{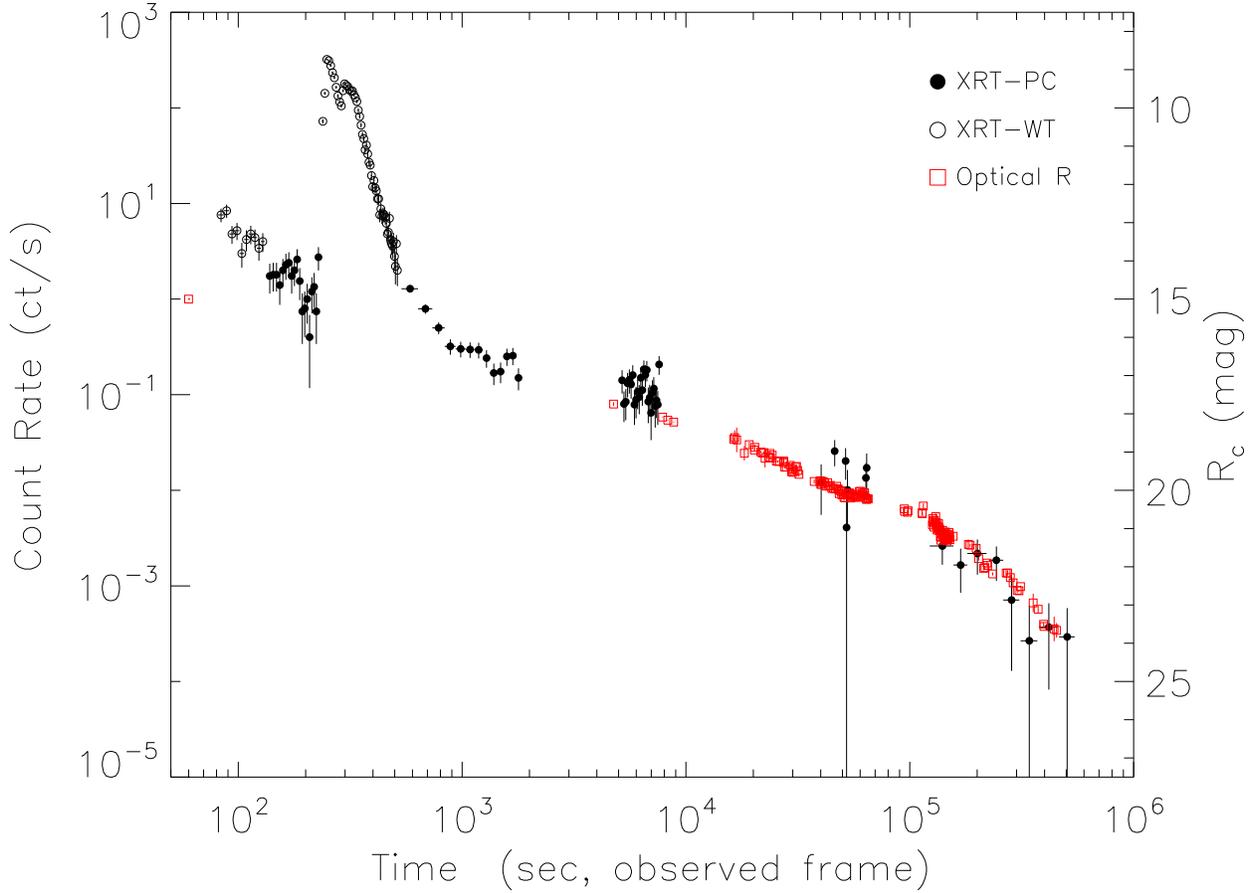}
\caption{X-ray and optical afterglow light curves of GRB~060526.  The open and filled circles are data from the XRT Windowed Timing mode and Photon Counting mode, respectively, and the red squares are the R band optical data from our MDM 1.3m, 2.4m, and PROMPT observations combined with other optical observations reported in the GCN. 
The normalization between the optical and X-ray light curves is $f_o/f_x\sim1000$ for the late afterglow light curve ($T>3\times10^4$~sec), and the optical to X-ray spectral index is $\beta_{ox}\sim1.0$.
\label{fig:lc}}
\end{figure}

\begin{figure}
\plotone{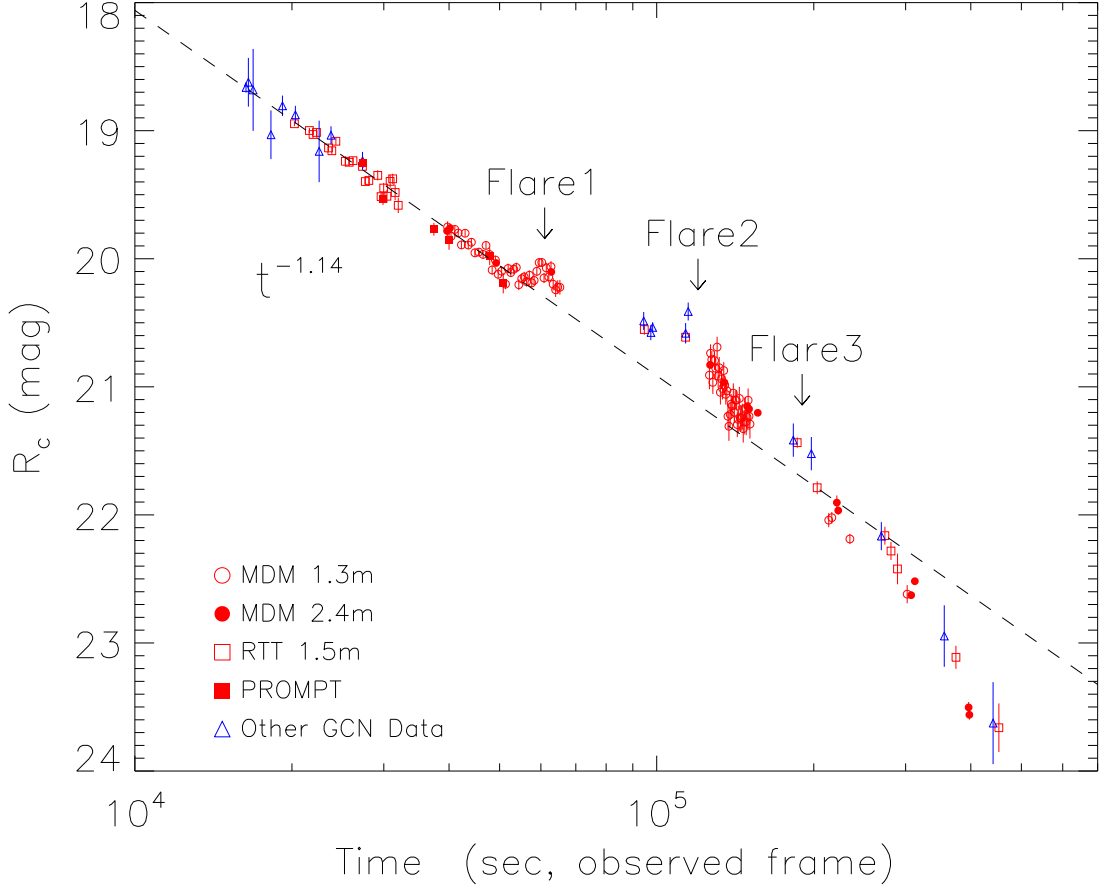}
\caption{The densely sampled optical light curve from $10^4$~sec after the burst trigger.  The open circles, filled circles, and fill squares are our MDM 1.3m, 2.4m, and PROMPT observations and the rest of the data are from GCN circulars.  
The data before $5.4\times10^4$~sec is fitted by a power-law with a slope of $1.14\pm0.02$ (dashed line).  Compared with the single power-law fit, the late optical data clearly show multiple flares and a much steeper late time decay slope of $3.4\pm0.2$.\label{fig:opt}}
\end{figure}

The overall optical and X-ray light-curves of GRB 060526 are shown in Fig.~\ref{fig:lc},
and data are listed in Tables~1 and 2. 
In addition, we also show the densely sampled optical data in more detail in Fig.~\ref{fig:opt}.  Both the optical and X-ray light curves exhibit rich features including flares and breaks beyond a simple power-law decay.  Below, we discuss the evolution of the optical and X-ray light curves separately at first and then compare them afterward.  Following the literature, we model the segments of the afterglow with single power-laws of $f_{\nu} \propto t^{-\alpha}\nu^{-\beta}$.  We also used the broken power-law for the temporal decay in some cases as $f(t) \propto ((\frac{t}{t_b})^{\alpha_es}+(\frac{t}{t_b})^{\alpha_ls})^{-1/s}$, where $\alpha_e$, $\alpha_l$, $t_b$, and $s$ are the early and late decay indices, break time, and smoothness parameters, respectively.  We fixed the smoothness parameter to $s=2.5$ for our analysis.

\begin{figure}
\epsscale{1}
\plotone{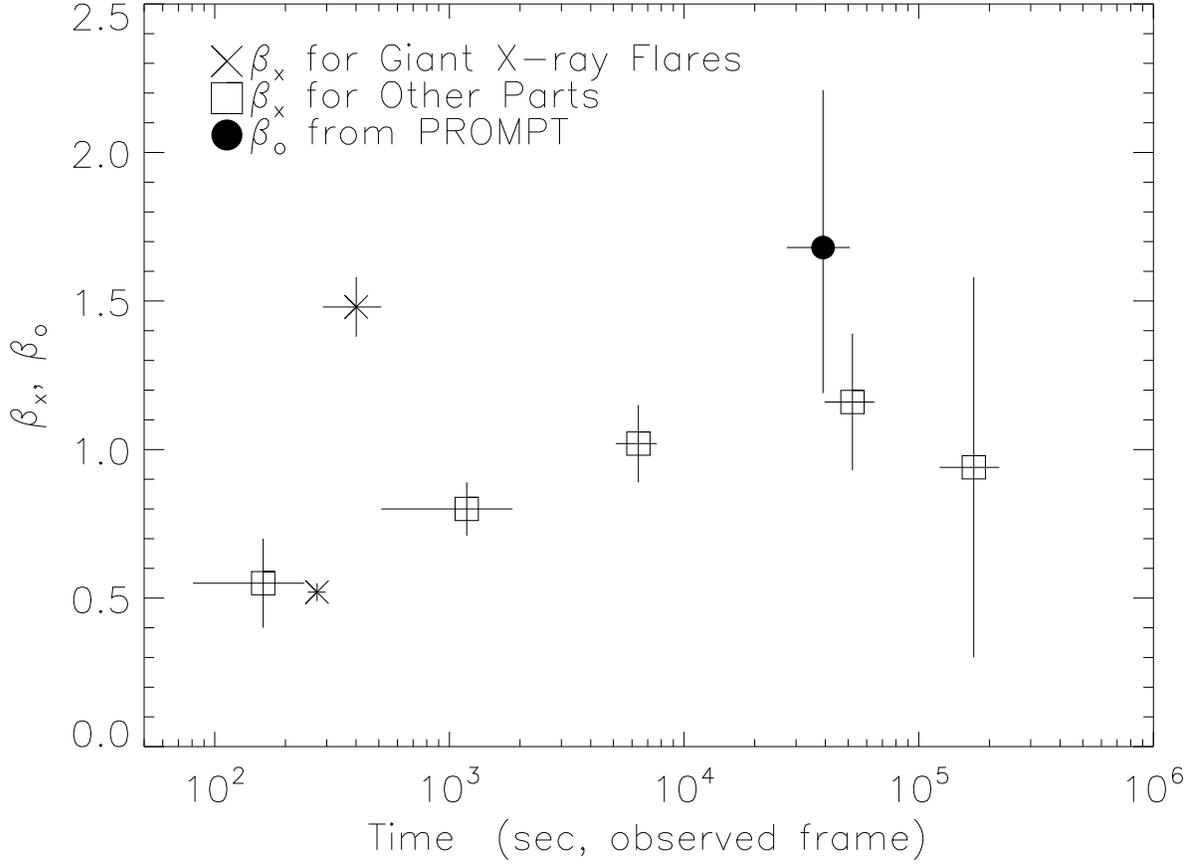}
\caption{The evolution of X-ray and optical afterglow spectral indices ($\beta_X$ and $\beta_o$) of GRB~060526.  The ``X'' symbols are the X-ray spectral indices for the two giant X-ray flares, and the squares are for other parts of the X-ray afterglow.  The filled circle is the optical spectral index obtained from PROMPT's $B, r', i'$ bands.\label{fig:spec}}
\end{figure}

The optical light curve shows at least one break at $T\sim2.4\times10^5$~sec and possibly an earlier one at $T\sim10^4$~sec.
In addition, we detect multiple optical flares in our densely sampled regions.  
The light curve between $10^4 <T<5.4\times10^4$~sec is well fitted by a power-law with a slope of $1.14\pm0.02$ 
(Fig.~\ref{fig:opt}).
In the period between $5.4\times10^4 <T<2.0\times10^5$~sec, several optical flares occurred, based on the shape of the light curve and the extrapolation of the power-law decay slope from the previous stage.
The flares peak at $\sim 6.1\times10^4$, $1.2\times10^5$, and $1.8\times10^5$~sec with peak magnitude changes of $\Delta m \sim 0.2, 0.6$, and $0.2$~mag, respectively, above a power-law extrapolation from the previous period,
and the durations of flares, $\Delta t$, are $1.4\pm0.5\times10^4$~sec, $9\pm2\times10^4$~sec, and $4\pm2\times10^4$~sec.
After the flares, the light curve steepens ($T>2.4\times10^5$~sec) as a power-law, and we find an index of $3.4\pm0.2$, 
assuming that the optical flares do not contribute to this part of the light curve.
If we treat the optical afterglow light curve as the ``bump and wiggle'' modification of the smooth afterglow instead of flares on top of a smooth afterglow, we fit all the optical data points ($T>10^4$~sec) with a broken power-law and we find early and late decay slopes of $\sim1.0$ and $\sim2.9$ and a break time at $\sim2.2\times10^5$~sec.  This fit is statistically unacceptable 
 because it cannot describe the ``bump and wiggle'' part of the afterglow.
We estimated the optical spectral index from the $B, r', i'$ images taken by the PROMPT close to $4\times10^4$~sec and obtained $\beta_o = 1.69^{+0.53}_{-0.49}$ (Fig.~\ref{fig:spec}).
We have taken account of the effects from Galactic extinction and the Ly$\alpha$ forest.

The most striking features of the X-ray light curve are the two X-ray flares between 225 and 600~sec after the burst trigger.
We find that the flares peak at $T=250$~sec and $T=300$~sec after the trigger, consistent with the GCN report \citep{campana06b}.  
If we neglect the data points dominated by the two X-ray flares, 
the light curve can be fitted with several power-law segments,
as with many other XRT afterglow light curves observed with \swift\ \citep{nousek06} and theoretical models \citep{zhang06}.
Besides the breaks and the two giant flares, the X-ray light curve also shows small amplitude variability between $5000 < T < 8000$~sec and possibly between $4\times10^4<T<6.4\times10^4$~sec.
The XRT light curve decays as $\alpha_1=2.3\pm0.2$ at $T<225$~sec before the flares.  
After the two X-ray flares, the X-ray emission decays slowly with $\alpha_2\sim0.5$ between $850 < T < 8000$~sec.
The extrapolation of the shallow decay slope does not fit the remaining data point, and the rest
of the afterglow evolution is expected to follow the ``normal+jet'' decay stage. 
However, the exact time of the transition is unknown because of the gaps in the XRT light curve.
If we assume the transition occurs at between $5000 < T <8000$~sec and fit the data points at $T>5000$~sec with a broken power-law model, we obtained $\alpha_3\sim1.1$, $\alpha_4\sim3.3$, and $T_b\sim2.3\times10^5$~sec.
We also fitted the X-ray light curve after the two giant flares ($T>850$~sec) using a single break and obtained a good fit with $\alpha_2\sim0.5$, $\alpha_3\sim1.6$, and $T_b\sim2\times10^4$~sec. 
Essentially, this is equivalent to replacing the previous ``normal+jet'' model with a single power-law decay of slope 1.6.
Finally, we analyzed the X-ray spectra of the GRB afterglow for several stages and show the evolution in Fig.~\ref{fig:spec}.
We detected spectral evolution before and after the two X-ray flares.
After the X-ray flares, we found the X-ray spectral indices are roughly consistent except that the index for the shallow decay stage ($\alpha_2 \sim 0.5$) is slightly harder. 

The late optical light curve ($T>10^4$~sec) clearly shows a break at $2.4\times10^5$~sec, while the X-ray light curve can be modeled both as a broken power-law and single power-law.
If the X-ray light curve decays as a single power-law at this stage, it obviously does not follow the optical afterglow.  On the other hand, the fitting results of the broken power-law model for the X-ray light curve are similar to those obtained from optical data.
Given the signal-to-noise of the late X-ray data and possible contamination from flares, it is difficult to distinguish the two models from the X-ray data alone.
Instead, since the properties of the optical afterglow are better constrained, we raise the question whether the X-ray afterglow is consistent with the best fit optical model.
The fitting results show that they are consistent, i.e., the optical and X-ray data are consistent with an achromatic broken power-law model with superimposed flares.

\section{Discussion\label{sec:dis}}
We present well-sampled optical and X-ray afterglow light curves of GRB~060526.  
As discussed in the previous section, the evolution of the afterglow is complicated with multiple breaks and flares both in the optical and the X-ray bands.  
The combination of flares and incomplete data sampling present severe challenges to measuring the temporal decay slopes of the afterglow, even for a well-sampled burst such as GRB~060526.
Below, we proceed by assuming that our analysis results are not significantly affected by these factors.

We detected a possible achromatic jet break
in the optical and X-ray afterglow light curves.
Before this late-time break ($T\sim2.4\times10^5$~sec), the afterglow is consistent with many \swift\ afterglows \citep{nousek06}.
The X-ray light curve started with a steep decay ($\alpha_1 = 2.3\pm0.2$), which is interpreted as the tail of the prompt emission due to the ``curvature effect'' \citep{kumar00}.  The spectral and temporal indices are constrained as $\alpha=\beta+2$, which is consistent with the X-ray spectral index of $\beta_1 = 0.55\pm0.15$.  The X-ray light curve then entered into a shallow decay stage with $\alpha_2 \sim 0.5$, which we interpret as energy injection.  Then it decays into a normal afterglow stage with $\alpha_3 = 1.14\pm0.02$ as constrained from the optical observations.
In addition, the X-ray light curve shows two huge flares which are commonly seen in \swift\ X-ray afterglows and are attributed to late time central engine activities.  After the achromatic break, the afterglow enters a very steep stage with $\alpha_4=3.4\pm0.2$.  
The X-ray and optical-to-X-ray spectral indices after $5000$~sec are consistent with $\sim 1$, suggesting that the optical and X-ray bands are on the same power-law segment of the spectral energy distribution.  
The optical spectral index is marginally consistent with the X-ray index, although the error-bar is large.

The achromatic break observed in optical afterglows is traditionally interpreted as the jet break.
As mentioned in the introduction, the achromatic break is not yet confirmed across optical and X-ray light curves.  Here, we present a case in GRB~060526 where such an achromatic break is observed across the optical and X-ray afterglows, supporting the beaming model of the GRBs.  However, the afterglow decay slopes before and after the break are hard to reconcile with simple afterglow-jet models.  The late time decay slope after the jet break should follow $\alpha = p$ \citep{sari99}, where $p$ is power-law index for the electrons $N(\gamma) \propto \gamma^{-p}$.  The post-jet slope, $3.4\pm0.2$, is too steep for a pre-jet slope of $1.14\pm0.02$ under any combination of either constant or wind medium and relative positions between $\nu$, $\nu_m$, and $\nu_c$.  Since the achromatic break is most easily explained by a jet, it is possible that more complicated afterglow models are needed \citep[e.g.,][]{panaitescu06b} with non-standard micro-physical parameters.
Another possibility is that energy injection or flares continued contributing significant flux and significantly affected the temporal decay slope.
We estimated the jet angle (half opening angle for uniform jets or observer's viewing angle for structured jets) using $t_j \simeq 6.2(E_{52}/n)^{1/3}(\theta_j/0.1)^{8/3}$~hr \citep{sari99} and obtained $\theta_j \sim 7^{\circ}$ assuming ambient density $n=1~$cm$^{-3}$.
We further estimated the size of the $\gamma$-ray prompt emission by combining the measured jet angle and the X-ray tail emission detected before 225~sec using $t_{tail} = (1+z)(R_{prompt}/c)(\theta_j^2/2)$ \citep{zhang06} and obtained $R_{prompt}\sim2\times10^{14}$~cm.

\begin{figure}
\plotone{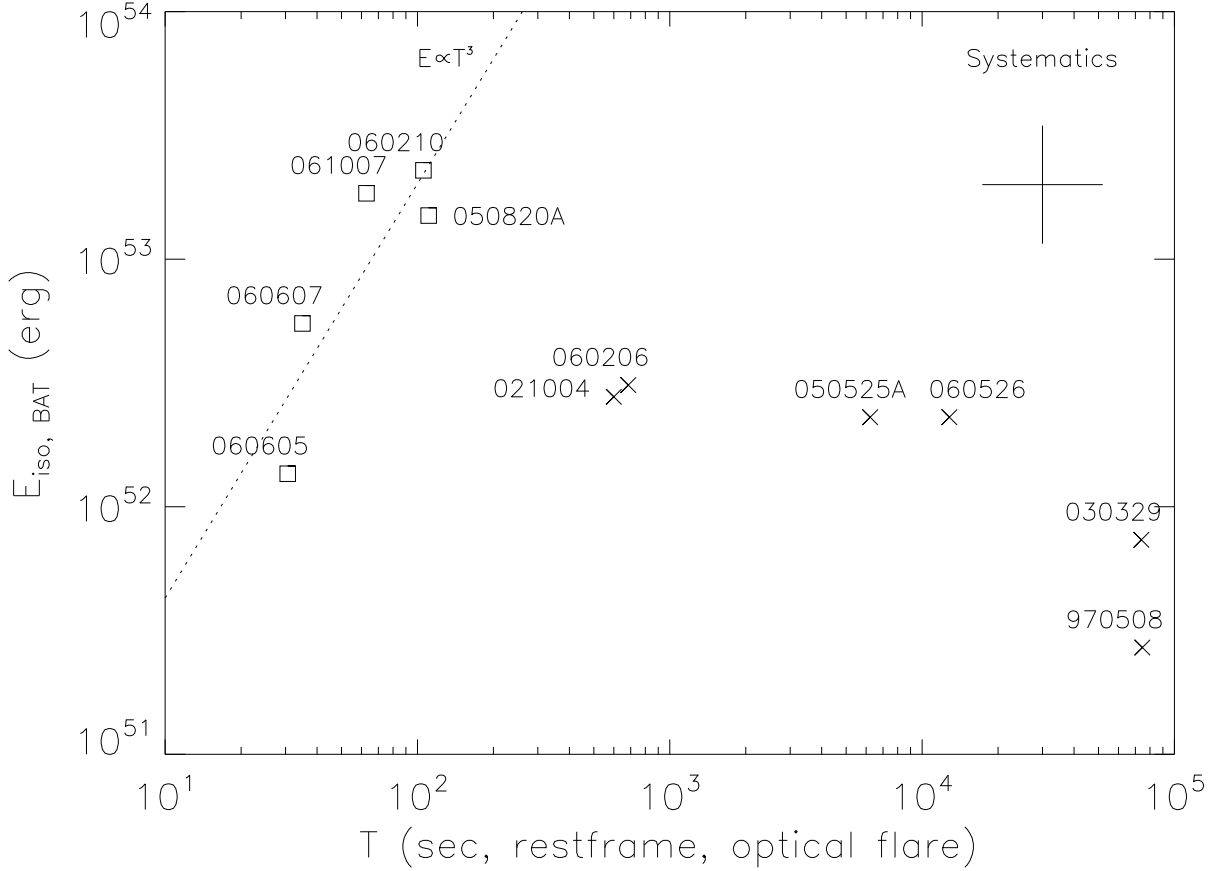} 
\caption{Isotropic energy in the BAT band versus optical flare time for GRBs with significant optical flares.  
We expect a factor of a few (3 in the plot as an example) for systematic uncertainties affecting the data points.  Overall, no positive correlation between isotropic energy and flare time is found from this data set.  However, a subset of the sample follows the prediction on the onset of the afterglow $T \propto E_{iso}^{1/3}$ (dotted line).
We notice the difference between the flares that occur before the afterglow has decayed (squares) and those afterward (``X'' symbols).  An important caveat is that the optical flare time is contingent on the initial optical observation of each event. \label{fig:et}}
\end{figure}

We also observed multiple optical flares in the light curve of GRB~060526.  Optical flares or re-brightenings have been observed in both pre-\swift\ \citep[e.g., GRB~970508, GRB~021004, and GRB~030329,][]{galama98,lazzati02,bersier03,mirabal03,lipkin04} and \swift\ bursts \citep[e.g., GRB~050525A, GRB~050820A, GRB~060206, GRB~060210, GRB~060605, GRB~060607, and GRB~061007,][]{blustin06,cenko06,stanek06,schaefer06,nysewander06,bersier06}.  
The fraction of bursts with optical flares seemed small.  However, recently many well-sampled bursts show complex optical decay behaviors.  The accumulating observations argue that it is possible that most of optical afterglows are complex and the appearance of simplicity was a consequence of poor sampling.
There are several interpretations for the optical flares, such as models of density fluctuations, ``patchy shell'', ``refreshed shock'', and late central engine activities \citep[e.g.,][]{jakobsson04,ioka05,gorosabel06}.
We estimated the quantities $\Delta t/t = 0.23\pm0.08$, $0.75\pm0.17$, and $0.22\pm0.11$ and $\Delta F_{\nu}/F_{\nu} \sim 0.2$, 0.7, and 0.2, respectively, for the three flares detected in GRB~060526.
The $\Delta t/t$ values are small which do not favor the models of patchy shell and refreshed shock, since the model predictions are $\Delta t/t > 1$ and $>1/4$ for these two models \citep{ioka05}.
The properties of the flares barely satisfy Ioka et al.'s constraint, $\Delta F_{\nu}/F_{\nu} < 1.6 \Delta t/t$, under the density fluctuation model.
Recently, \citet{nakar06} also modeled the effects of density fluctuations on the afterglow light curves and found that they cannot produce the sharp features observed in many bursts.
Another possibility is that the flares (or breaks) indicate the onset of the afterglow \citep{rykoff06a,stanek06} scaled with the isotropic energy as $T_{onset} \propto E_{iso}^{1/3}$ \citep{sari97}.
The onset time also depends on the density of the ambient medium and the initial Lorentz factor that are more difficult to measure.
We might expect a correlation between $E_{iso}$ and $T_{onset}$ for a large sample of bursts, or if the densities and Lorentz factors for the bursts only spread in a narrow range.
We tested this hypothesis by plotting the two properties for bursts with optical flares in Fig.~\ref{fig:et}, and did not detect positive correlation between the two properties.  
However, we notice the difference between flares that occur before the optical afterglow has decayed and those that occur afterward.
The flares in GRB~050820A, GRB~060210, GRB~060605, GRB~060607, and GRB~061007 possibly belong to the category of flares that occur before the afterglow has faded, and they roughly follow the scaling between isotropic energy and flare time.
However, a larger sample is needed to fully test the model.
In any case, the flares in GRB~060526 are unlikely to be associated with the onset of the afterglow.
It is possible that these flares are from late central engine activities, which can have arbitrary variabilities.  However, we are open to other theoretical models which can be tested extensively with our well-sampled light curve.  

\acknowledgements 
We acknowledge the \swift\ team for the prompt detection and localization of the GRB and the rapid release of data products.  We also thank the GRB Coordinates Network (GCN) and astronomers who contribute to the GCN circular.  We thank B. Zhang for helpful discussion.

\clearpage

\begin{deluxetable}{cccc}
\tabletypesize{\scriptsize}
\tablecolumns{4}
\tablewidth{0pt}
\tablecaption{Optical Light Curves of GRB~060526. \label{tab:datao}}
\tablehead{
\colhead{Telescope} &
\colhead{Time} &
\colhead{Band} &
\colhead{Magnitude} \\
\colhead{} &
\colhead{(sec)} &
\colhead{} &
\colhead{} 
}

\startdata
MDM 1.3m & 39744 & R  & 19.75$\pm$0.04  \\
MDM 1.3m & 40349 & R  & 19.83$\pm$0.04  \\
MDM 1.3m & 41040 & R  & 19.77$\pm$0.03  \\
MDM 1.3m & 41645 & R  & 19.80$\pm$0.03  \\
MDM 1.3m & 42250 & R  & 19.89$\pm$0.03  \\
MDM 1.3m & 42941 & R  & 19.80$\pm$0.03  \\
MDM 1.3m & 43546 & R  & 19.89$\pm$0.03  \\
MDM 1.3m & 44150 & R  & 19.87$\pm$0.03  \\
MDM 1.3m & 44842 & R  & 19.95$\pm$0.04  \\
MDM 1.3m & 45619 & R  & 19.95$\pm$0.03  \\
MDM 1.3m & 46483 & R  & 19.97$\pm$0.03  \\
MDM 1.3m & 47088 & R  & 19.90$\pm$0.03  \\
MDM 1.3m & 47779 & R  & 19.97$\pm$0.03  \\
MDM 1.3m & 48384 & R  & 20.09$\pm$0.04  \\
MDM 1.3m & 49075 & R  & 20.01$\pm$0.03  \\
MDM 1.3m & 49680 & R  & 20.12$\pm$0.04  \\
MDM 1.3m & 50458 & R  & 20.10$\pm$0.04  \\
MDM 1.3m & 51322 & R  & 20.20$\pm$0.04  \\
MDM 1.3m & 51926 & R  & 20.08$\pm$0.04  \\
MDM 1.3m & 52531 & R  & 20.11$\pm$0.04  \\
MDM 1.3m & 53222 & R  & 20.08$\pm$0.04  \\
MDM 1.3m & 53827 & R  & 20.07$\pm$0.04  \\
MDM 1.3m & 54432 & R  & 20.20$\pm$0.04  \\
MDM 1.3m & 55123 & R  & 20.16$\pm$0.04  \\
MDM 1.3m & 55728 & R  & 20.14$\pm$0.04  \\
MDM 1.3m & 56333 & R  & 20.18$\pm$0.04  \\
MDM 1.3m & 57024 & R  & 20.13$\pm$0.04  \\
MDM 1.3m & 57629 & R  & 20.19$\pm$0.04  \\
MDM 1.3m & 58234 & R  & 20.17$\pm$0.04  \\
MDM 1.3m & 58925 & R  & 20.10$\pm$0.04  \\
MDM 1.3m & 59530 & R  & 20.03$\pm$0.04  \\
MDM 1.3m & 60221 & R  & 20.03$\pm$0.04  \\
MDM 1.3m & 60826 & R  & 20.15$\pm$0.04  \\
MDM 1.3m & 61430 & R  & 20.07$\pm$0.04  \\
MDM 1.3m & 62122 & R  & 20.14$\pm$0.04  \\
MDM 1.3m & 62726 & R  & 20.06$\pm$0.04  \\
MDM 1.3m & 63331 & R  & 20.20$\pm$0.05  \\
MDM 1.3m & 64022 & R  & 20.24$\pm$0.06  \\
MDM 1.3m & 64627 & R  & 20.22$\pm$0.05  \\
MDM 1.3m & 65232 & R  & 20.22$\pm$0.06  \\
MDM 1.3m & 126230 & R  & 20.91$\pm$0.11  \\
MDM 1.3m & 126835 & R  & 20.74$\pm$0.07  \\
MDM 1.3m & 127526 & R  & 20.79$\pm$0.07  \\
MDM 1.3m & 128131 & R  & 20.97$\pm$0.09  \\
MDM 1.3m & 129082 & R  & 20.80$\pm$0.07  \\
MDM 1.3m & 129773 & R  & 20.86$\pm$0.08  \\
MDM 1.3m & 130550 & R  & 20.69$\pm$0.08  \\
MDM 1.3m & 131155 & R  & 20.92$\pm$0.09  \\
MDM 1.3m & 131760 & R  & 20.85$\pm$0.08  \\
MDM 1.3m & 132451 & R  & 21.04$\pm$0.10  \\
MDM 1.3m & 133142 & R  & 20.94$\pm$0.08  \\
MDM 1.3m & 133747 & R  & 21.01$\pm$0.09  \\
MDM 1.3m & 134438 & R  & 20.87$\pm$0.07  \\
MDM 1.3m & 135043 & R  & 20.99$\pm$0.07  \\
MDM 1.3m & 135648 & R  & 21.06$\pm$0.08  \\
MDM 1.3m & 136339 & R  & 21.03$\pm$0.09  \\
MDM 1.3m & 136944 & R  & 21.23$\pm$0.10  \\
MDM 1.3m & 137549 & R  & 21.31$\pm$0.11  \\
MDM 1.3m & 138326 & R  & 21.22$\pm$0.09  \\
MDM 1.3m & 138931 & R  & 21.14$\pm$0.08  \\
MDM 1.3m & 139536 & R  & 21.15$\pm$0.08  \\
MDM 1.3m & 140227 & R  & 21.05$\pm$0.08  \\
MDM 1.3m & 140832 & R  & 21.20$\pm$0.09  \\
MDM 1.3m & 141437 & R  & 21.10$\pm$0.07  \\
MDM 1.3m & 142128 & R  & 21.11$\pm$0.08  \\
MDM 1.3m & 142733 & R  & 21.30$\pm$0.10  \\
MDM 1.3m & 143338 & R  & 21.25$\pm$0.11  \\
MDM 1.3m & 144029 & R  & 21.09$\pm$0.09  \\
MDM 1.3m & 144634 & R  & 21.25$\pm$0.10  \\
MDM 1.3m & 145238 & R  & 21.24$\pm$0.10  \\
MDM 1.3m & 145930 & R  & 21.23$\pm$0.10  \\
MDM 1.3m & 146534 & R  & 21.33$\pm$0.11  \\
MDM 1.3m & 147139 & R  & 21.28$\pm$0.10  \\
MDM 1.3m & 147830 & R  & 21.17$\pm$0.09  \\
MDM 1.3m & 148435 & R  & 21.27$\pm$0.10  \\
MDM 1.3m & 149040 & R  & 21.24$\pm$0.10  \\
MDM 1.3m & 149731 & R  & 21.11$\pm$0.09  \\
MDM 1.3m & 150336 & R  & 21.23$\pm$0.11  \\
MDM 1.3m & 150941 & R  & 21.29$\pm$0.11  \\
MDM 1.3m & 213667 & R  & 22.04$\pm$0.06  \\
MDM 1.3m & 216605 & R  & 22.02$\pm$0.05  \\
MDM 1.3m & 234576 & R  & 22.19$\pm$0.04  \\
MDM 1.3m & 302054 & R  & 22.62$\pm$0.07  \\
MDM 2.4m & 39658 & R  & 19.78$\pm$0.03  \\
MDM 2.4m & 40176 & R  & 19.76$\pm$0.02  \\
MDM 2.4m & 49248 & R  & 20.03$\pm$0.02  \\
MDM 2.4m & 62813 & R  & 20.10$\pm$0.03  \\
MDM 2.4m & 126576 & R  & 20.83$\pm$0.03  \\
MDM 2.4m & 134784 & R  & 20.97$\pm$0.02  \\
MDM 2.4m & 149040 & R  & 21.15$\pm$0.03  \\
MDM 2.4m & 150336 & R  & 21.17$\pm$0.02  \\
MDM 2.4m & 156211 & R  & 21.20$\pm$0.03  \\
MDM 2.4m & 221443 & R  & 21.90$\pm$0.06  \\
MDM 2.4m & 222826 & R  & 21.97$\pm$0.03  \\
MDM 2.4m & 307411 & R  & 22.63$\pm$0.03  \\
MDM 2.4m & 312422 & R  & 22.52$\pm$0.02  \\
MDM 2.4m & 396144 & R  & 23.50$\pm$0.04  \\
MDM 2.4m & 397440 & R  & 23.56$\pm$0.04  \\
PROMPT & 27000 & B  & 20.73$\pm$0.08  \\
PROMPT & 36000 & B  & 21.36$\pm$0.09  \\
PROMPT & 43560 & B  & 21.33$\pm$0.09  \\
PROMPT & 48960 & B  & 21.85$\pm$0.15  \\
PROMPT & 55440 & B  & 21.94$\pm$0.22  \\
PROMPT & 27360 & r'  & 19.49$\pm$0.04  \\
PROMPT & 29880 & r'  & 19.77$\pm$0.05  \\
PROMPT & 37440 & r'  & 20.01$\pm$0.05  \\
PROMPT & 39960 & r'  & 20.09$\pm$0.08  \\
PROMPT & 47880 & r'  & 20.22$\pm$0.07  \\
PROMPT & 50760 & r'  & 20.43$\pm$0.08  \\
PROMPT & 24840 & i'  & 19.22$\pm$0.06  \\
PROMPT & 32760 & i'  & 19.56$\pm$0.06  \\
PROMPT & 35280 & i'  & 19.77$\pm$0.06  \\
PROMPT & 42480 & i'  & 19.79$\pm$0.06  \\
PROMPT & 45000 & i'  & 19.97$\pm$0.04  \\
PROMPT & 53640 & i'  & 20.00$\pm$0.08  \\
PROMPT & 57240 & i'  & 19.98$\pm$0.10  \\
\enddata

\end{deluxetable}

\clearpage

\begin{deluxetable}{ccccc}
\tabletypesize{\scriptsize}
\tablecolumns{5}
\tablewidth{0pt}
\tablecaption{X-ray Light Curves of GRB~060526. \label{tab:datax}}
\tablehead{
\colhead{Telescope} &
\colhead{Time} &
\colhead{Band} &
\colhead{Mode} &
\colhead{Count Rate} \\
\colhead{} &
\colhead{(sec)} &
\colhead{} &
\colhead{} &
\colhead{(count~s$^{-1}$)}
}

\startdata
\swift & 84 & 0.2--10 keV & WT  & 7.60$\pm$1.23 \\
\swift & 89 & 0.2--10 keV & WT  & 8.40$\pm$1.33 \\
\swift & 94 & 0.2--10 keV & WT  & 4.80$\pm$1.02 \\
\swift & 99 & 0.2--10 keV & WT  & 5.20$\pm$1.06 \\
\swift & 104 & 0.2--10 keV & WT  & 3.00$\pm$0.87 \\
\swift & 109 & 0.2--10 keV & WT  & 4.20$\pm$1.04 \\
\swift & 114 & 0.2--10 keV & WT  & 4.80$\pm$1.06 \\
\swift & 119 & 0.2--10 keV & WT  & 4.40$\pm$0.94 \\
\swift & 124 & 0.2--10 keV & WT  & 3.40$\pm$0.87 \\
\swift & 129 & 0.2--10 keV & WT  & 4.00$\pm$0.89 \\
\swift & 239 & 0.2--10 keV & WT  & 72.40$\pm$3.83 \\
\swift & 244 & 0.2--10 keV & WT  & 142.20$\pm$5.36 \\
\swift & 249 & 0.2--10 keV & WT  & 321.80$\pm$8.06 \\
\swift & 254 & 0.2--10 keV & WT  & 312.00$\pm$7.96 \\
\swift & 259 & 0.2--10 keV & WT  & 276.80$\pm$7.50 \\
\swift & 264 & 0.2--10 keV & WT  & 233.60$\pm$6.90 \\
\swift & 269 & 0.2--10 keV & WT  & 206.60$\pm$6.43 \\
\swift & 274 & 0.2--10 keV & WT  & 164.40$\pm$5.80 \\
\swift & 279 & 0.2--10 keV & WT  & 133.60$\pm$5.20 \\
\swift & 284 & 0.2--10 keV & WT  & 115.60$\pm$4.83 \\
\swift & 289 & 0.2--10 keV & WT  & 105.00$\pm$4.59 \\
\swift & 294 & 0.2--10 keV & WT  & 151.20$\pm$5.54 \\
\swift & 299 & 0.2--10 keV & WT  & 178.60$\pm$6.01 \\
\swift & 304 & 0.2--10 keV & WT  & 172.00$\pm$5.89 \\
\swift & 309 & 0.2--10 keV & WT  & 167.00$\pm$5.79 \\
\swift & 314 & 0.2--10 keV & WT  & 155.80$\pm$5.63 \\
\swift & 319 & 0.2--10 keV & WT  & 149.80$\pm$5.49 \\
\swift & 324 & 0.2--10 keV & WT  & 150.80$\pm$5.51 \\
\swift & 329 & 0.2--10 keV & WT  & 136.80$\pm$5.24 \\
\swift & 334 & 0.2--10 keV & WT  & 128.00$\pm$5.11 \\
\swift & 339 & 0.2--10 keV & WT  & 116.80$\pm$4.87 \\
\swift & 344 & 0.2--10 keV & WT  & 94.60$\pm$4.35 \\
\swift & 349 & 0.2--10 keV & WT  & 81.40$\pm$4.03 \\
\swift & 354 & 0.2--10 keV & WT  & 66.00$\pm$3.67 \\
\swift & 359 & 0.2--10 keV & WT  & 52.80$\pm$3.27 \\
\swift & 364 & 0.2--10 keV & WT  & 47.60$\pm$3.11 \\
\swift & 369 & 0.2--10 keV & WT  & 36.40$\pm$2.73 \\
\swift & 374 & 0.2--10 keV & WT  & 40.80$\pm$2.88 \\
\swift & 379 & 0.2--10 keV & WT  & 33.00$\pm$2.62 \\
\swift & 384 & 0.2--10 keV & WT  & 27.00$\pm$2.34 \\
\swift & 389 & 0.2--10 keV & WT  & 25.20$\pm$2.28 \\
\swift & 394 & 0.2--10 keV & WT  & 19.60$\pm$2.00 \\
\swift & 399 & 0.2--10 keV & WT  & 15.00$\pm$1.78 \\
\swift & 404 & 0.2--10 keV & WT  & 17.40$\pm$1.89 \\
\swift & 409 & 0.2--10 keV & WT  & 14.60$\pm$1.71 \\
\swift & 414 & 0.2--10 keV & WT  & 13.60$\pm$1.67 \\
\swift & 419 & 0.2--10 keV & WT  & 11.20$\pm$1.57 \\
\swift & 424 & 0.2--10 keV & WT  & 11.20$\pm$1.50 \\
\swift & 429 & 0.2--10 keV & WT  & 7.60$\pm$1.23 \\
\swift & 434 & 0.2--10 keV & WT  & 8.80$\pm$1.33 \\
\swift & 439 & 0.2--10 keV & WT  & 7.40$\pm$1.28 \\
\swift & 444 & 0.2--10 keV & WT  & 7.80$\pm$1.25 \\
\swift & 449 & 0.2--10 keV & WT  & 7.60$\pm$1.30 \\
\swift & 454 & 0.2--10 keV & WT  & 7.20$\pm$1.20 \\
\swift & 459 & 0.2--10 keV & WT  & 6.20$\pm$1.18 \\
\swift & 464 & 0.2--10 keV & WT  & 4.80$\pm$1.02 \\
\swift & 469 & 0.2--10 keV & WT  & 5.00$\pm$1.00 \\
\swift & 474 & 0.2--10 keV & WT  & 7.00$\pm$1.22 \\
\swift & 479 & 0.2--10 keV & WT  & 4.20$\pm$0.96 \\
\swift & 484 & 0.2--10 keV & WT  & 3.80$\pm$0.87 \\
\swift & 489 & 0.2--10 keV & WT  & 3.60$\pm$0.94 \\
\swift & 494 & 0.2--10 keV & WT  & 4.00$\pm$0.94 \\
\swift & 499 & 0.2--10 keV & WT  & 2.80$\pm$0.75 \\
\swift & 504 & 0.2--10 keV & WT  & 2.20$\pm$0.77 \\
\swift & 509 & 0.2--10 keV & WT  & 3.80$\pm$0.87 \\
\swift & 514 & 0.2--10 keV & WT  & 2.00$\pm$0.63 \\
\swift & 138 & 0.2--10 keV & PC  & 1.7436$\pm$0.6026 \\
\swift & 143 & 0.2--10 keV & PC  & 1.8000$\pm$0.6000 \\
\swift & 148 & 0.2--10 keV & PC  & 1.8000$\pm$0.6000 \\
\swift & 153 & 0.2--10 keV & PC  & 1.4000$\pm$0.5292 \\
\swift & 158 & 0.2--10 keV & PC  & 2.0000$\pm$0.6325 \\
\swift & 163 & 0.2--10 keV & PC  & 2.2872$\pm$0.6974 \\
\swift & 168 & 0.2--10 keV & PC  & 2.4000$\pm$0.6928 \\
\swift & 173 & 0.2--10 keV & PC  & 1.7436$\pm$0.6026 \\
\swift & 178 & 0.2--10 keV & PC  & 2.0000$\pm$0.6325 \\
\swift & 183 & 0.2--10 keV & PC  & 2.6000$\pm$0.7211 \\
\swift & 188 & 0.2--10 keV & PC  & 1.5436$\pm$0.5685 \\
\swift & 193 & 0.2--10 keV & PC  & 0.7436$\pm$0.4040 \\
\swift & 198 & 0.2--10 keV & PC  & 0.8000$\pm$0.4000 \\
\swift & 203 & 0.2--10 keV & PC  & 1.0000$\pm$0.4472 \\
\swift & 208 & 0.2--10 keV & PC  & 0.4000$\pm$0.2828 \\
\swift & 213 & 0.2--10 keV & PC  & 1.2000$\pm$0.4899 \\
\swift & 218 & 0.2--10 keV & PC  & 1.3436$\pm$0.5321 \\
\swift & 223 & 0.2--10 keV & PC  & 0.7436$\pm$0.4040 \\
\swift & 228 & 0.2--10 keV & PC  & 2.7436$\pm$0.7505 \\
\swift & 586 & 0.2--10 keV & PC  & 1.2844$\pm$0.1136 \\
\swift & 686 & 0.2--10 keV & PC  & 0.7915$\pm$0.0896 \\
\swift & 786 & 0.2--10 keV & PC  & 0.5000$\pm$0.0707 \\
\swift & 886 & 0.2--10 keV & PC  & 0.3203$\pm$0.0588 \\
\swift & 986 & 0.2--10 keV & PC  & 0.3015$\pm$0.0559 \\
\swift & 1086 & 0.2--10 keV & PC  & 0.2972$\pm$0.0548 \\
\swift & 1186 & 0.2--10 keV & PC  & 0.2944$\pm$0.0549 \\
\swift & 1286 & 0.2--10 keV & PC  & 0.2415$\pm$0.0502 \\
\swift & 1386 & 0.2--10 keV & PC  & 0.1687$\pm$0.0428 \\
\swift & 1486 & 0.2--10 keV & PC  & 0.1744$\pm$0.0426 \\
\swift & 1586 & 0.2--10 keV & PC  & 0.2515$\pm$0.0512 \\
\swift & 1686 & 0.2--10 keV & PC  & 0.2559$\pm$0.0523 \\
\swift & 1786 & 0.2--10 keV & PC  & 0.1500$\pm$0.0387 \\
\swift & 5186 & 0.2--10 keV & PC  & 0.1415$\pm$0.0390 \\
\swift & 5286 & 0.2--10 keV & PC  & 0.0800$\pm$0.0283 \\
\swift & 5386 & 0.2--10 keV & PC  & 0.0844$\pm$0.0303 \\
\swift & 5486 & 0.2--10 keV & PC  & 0.1315$\pm$0.0377 \\
\swift & 5586 & 0.2--10 keV & PC  & 0.1431$\pm$0.0406 \\
\swift & 5686 & 0.2--10 keV & PC  & 0.1287$\pm$0.0378 \\
\swift & 5786 & 0.2--10 keV & PC  & 0.1603$\pm$0.0431 \\
\swift & 5886 & 0.2--10 keV & PC  & 0.0787$\pm$0.0305 \\
\swift & 5986 & 0.2--10 keV & PC  & 0.0887$\pm$0.0321 \\
\swift & 6086 & 0.2--10 keV & PC  & 0.1074$\pm$0.0369 \\
\swift & 6186 & 0.2--10 keV & PC  & 0.0944$\pm$0.0319 \\
\swift & 6286 & 0.2--10 keV & PC  & 0.1503$\pm$0.0419 \\
\swift & 6386 & 0.2--10 keV & PC  & 0.1115$\pm$0.0350 \\
\swift & 6486 & 0.2--10 keV & PC  & 0.1844$\pm$0.0438 \\
\swift & 6586 & 0.2--10 keV & PC  & 0.1615$\pm$0.0415 \\
\swift & 6686 & 0.2--10 keV & PC  & 0.1815$\pm$0.0439 \\
\swift & 6786 & 0.2--10 keV & PC  & 0.0846$\pm$0.0342 \\
\swift & 6886 & 0.2--10 keV & PC  & 0.0931$\pm$0.0339 \\
\swift & 6986 & 0.2--10 keV & PC  & 0.0646$\pm$0.0312 \\
\swift & 7086 & 0.2--10 keV & PC  & 0.1046$\pm$0.0370 \\
\swift & 7186 & 0.2--10 keV & PC  & 0.1159$\pm$0.0366 \\
\swift & 7286 & 0.2--10 keV & PC  & 0.0759$\pm$0.0307 \\
\swift & 7386 & 0.2--10 keV & PC  & 0.0872$\pm$0.0301 \\
\swift & 7486 & 0.2--10 keV & PC  & 0.0787$\pm$0.0305 \\
\swift & 7586 & 0.2--10 keV & PC  & 0.2072$\pm$0.0459 \\
\swift & 40173 & 0.2--10 keV & PC  & 0.0121$\pm$0.0065 \\
\swift & 46173 & 0.2--10 keV & PC  & 0.0257$\pm$0.0078 \\
\swift & 51673 & 0.2--10 keV & PC  & 0.0202$\pm$0.0074 \\
\swift & 52173 & 0.2--10 keV & PC  & 0.0041$\pm$0.0052 \\
\swift & 52673 & 0.2--10 keV & PC  & 0.0101$\pm$0.0062 \\
\swift & 63673 & 0.2--10 keV & PC  & 0.0135$\pm$0.0057 \\
\swift & 64173 & 0.2--10 keV & PC  & 0.0172$\pm$0.0071 \\
\swift & 139785 & 0.2--10 keV & PC  & 0.0026$\pm$0.0009 \\
\swift & 168735 & 0.2--10 keV & PC  & 0.0017$\pm$0.0008 \\
\swift & 200435 & 0.2--10 keV & PC  & 0.0022$\pm$0.0009 \\
\swift & 243601 & 0.2--10 keV & PC  & 0.0019$\pm$0.0007 \\
\swift & 284976 & 0.2--10 keV & PC  & 0.0007$\pm$0.0006 \\
\swift & 341833 & 0.2--10 keV & PC  & 0.0003$\pm$0.0004 \\
\swift & 504585 & 0.2--10 keV & PC  & 0.0003$\pm$0.0003 \\
\swift & 417948 & 0.2--10 keV & PC  & 0.0004$\pm$0.0003 \\
\enddata

\end{deluxetable}

\clearpage

\clearpage

\clearpage

\clearpage

\end{document}